\documentclass[%
twocolumn,
superscriptaddress,
 amsmath,amssymb,
 aps,
prl,
noeprint,
noinfoline
]{revtex4-2}

\usepackage{graphicx}
\usepackage{dcolumn}
\usepackage{bm}
\usepackage{float}
\usepackage{verbatim}

\usepackage{color}

\begin{document}

\preprint{APS/123-QED}

\title{Resonance-enhanced Floquet cavity electromagnonics 
}

\author{Amin Pishehvar}
\altaffiliation{These authors contributed equally to this work.}
\affiliation{ 
    Department of Electrical and Computer Engineering, Northeastern University, Boston, MA 02115, USA
}

\author{Zixin Yan}
\altaffiliation{These authors contributed equally to this work.}
\affiliation{ 
    Department of Electrical and Computer Engineering, Northeastern University, Boston, MA 02115, USA
}

\author{Zhaoyou Wang}
\affiliation{ 
    Pritzker School of Molecular Engineering, University of Chicago, Chicago, IL 60637, USA
}

\author{Yu Jiang}
\affiliation{ 
    Department of Electrical and Computer Engineering, Northeastern University, Boston, MA 02115, USA
}

\author{Yizhong Huang}
\affiliation{ 
    Department of Electrical and Computer Engineering, Northeastern University, Boston, MA 02115, USA
}

\author{Josep M. Jornet}
\affiliation{ 
    Department of Electrical and Computer Engineering, Northeastern University, Boston, MA 02115, USA
}

\author{Liang Jiang}
\affiliation{ 
    Pritzker School of Molecular Engineering, University of Chicago, Chicago, IL 60637, USA
}

\author{Xufeng Zhang}
\email{xu.zhang@northeastern.edu}
\affiliation{ 
    Department of Electrical and Computer Engineering, Northeastern University, Boston, MA 02115, USA
}
\affiliation{ 
    Department of Physics, Northeastern University, Boston, MA 02115, USA
}

\date{\today}

\begin{abstract}
Floquet engineering has been recently recognized as an important tool for manipulating the coherent magnon-photon interaction in cavity electromagnonics systems at microwave frequencies. In spite of the novel hybrid magnonic functionalities that have been demonstrated, the effect of the Floquet drive has been relatively weak due to the limited driving efficiency, limiting its broader application. This work shows that by utilizing LC resonances, the Floquet drive in our cavity electromagnonic device can be drastically enhanced, giving rise to drastically boosted interaction between hybrid modes with fundamentally different spectral characteristics compared with previous demonstrations.  In addition, the Floquet drives can also be obtained from GHz signals on such a system, allowing the demonstration of more advanced signal operations. Our novel resonance-enhanced Floquet cavity electromagnonics points to a new direction to fully unleash the potential of Floquet hybrid magnonics.
\end{abstract}


\maketitle


\section{Introduction} 

Hybrid magnonics \cite{Li2020HybridMagnonics,Tabuchi2019HybridQuantum,Chumak2021QuantumEngineering,LachanceQuirion2022HybridQuantum,ZHANG2023100044,ZareRameshti2022Sep,HARDER201847,BHOI202039} has recently emerged as a novel and rapidly developing field of research. In this field, magnons -- elementary collective excitations of magnetization -- interact with other fundamental excitations such as photons and phonons \cite{PhysRevLett.113.156401,PhysRevLett.113.083603,PhysRevApplied.2.054002,PhysRevLett.114.227201,PhysRevLett.117.123605,Osada2016APS,PhysRevLett.117.133602,PhysRevA.94.033821,doi:10.1126/sciadv.1501286,PhysRevX.11.031053}. The hybridized modes (magnon polaritons) arising from these interactions inherit characteristics of magnons, including nonreciprocity \cite{PhysRevLett.123.127202,PhysRevApplied.13.044039} and nonlinearity \cite{Wang2024AllMagnonicRepeater,PhysRevLett.120.057202,PhysRevB.94.224410}, as well as properties of the other excitations. This combination allows the integration of diverse functionalities on a single platform, facilitating the discovery of novel physical phenomena and technological applications. For example, magnons can strongly couple to microwave photons within cavity electromagnonic systems, a highly versatile approach implemented across various physical platforms. These systems typically involve coupled magnonic resonators (such as spheres or thin films made of single-crystalline ferrimagnetic insulator yttrium iron garnet, i.e., YIG) integrated with microwave resonators or waveguides in diverse forms, including three-dimensional (3D) cavities \cite{PhysRevLett.113.083603,PhysRevLett.113.156401,PhysRevApplied.2.054002,PhysRevLett.114.227201,Braggio2017PRL}, dielectric resonators \cite{Xu2020Dec_PRL,Kato2023Aug}, 
planar resonators \cite{Huebl2013SepPRL,Bhoi2017Sep,PhysRevLett.123.107701,PhysRevLett.123.107702,PhysRevApplied.21.034034},
and spoof surface plasmon polariton waveguides \cite{PhysRevLett.132.116701,Jiang2025FebPRAppl} constructed from materials such as dielectric, normal metal, or superconductor. These hybrid systems have paved the way for innovative applications in quantum transduction \cite{doi:10.1126/science.aaa3693,doi:10.1126/sciadv.1603150,doi:10.1126/science.aaz9236,PhysRevLett.130.193603,PhysRevLett.125.117701,Zhu:20Optica,Haigh2015PRA,Osada2016APS,Zhang2016APS,Sivarajah2019JAP,Crescini2020MagnonHybrid,Rashedi2024YIGPhotonic,Haigh2020Subpicoliter}, neuromorphic computing \cite{Yaremkevich2023OnChipPhononMagnon,Millet2021APL}, dark matter detection \cite{FLOWER2019DarkUniverse,Crescini2020APL,Crescini2018AxionHaloscope}, and quantum sensing \cite{doi:10.1126/science.aaz9236,doi:10.1126/sciadv.1603150,PhysRevLett.125.117701,PhysRevLett.130.193603,Crescini2020APL,PhysRevApplied.13.064001}, and continue to expand rapidly.

Dynamic tunability of interactions within hybrid magnonic systems is essential for unlocking new physics and applications. Recently, Floquet engineering has been recognized as an effective method to dynamically manipulate magnon-photon interactions at microwave frequencies, leading to the establishment of Floquet cavity electromagnonics \cite{Xu2020Dec_PRL,Pishehvar2025Feb_PRAppl}. In these systems, hybrid modes are generated from the coherent magnon-photon coupling. When a Floquet drive is applied to the system with a frequency matching the energy spacing between these two hybrid modes, each hybrid mode splits into two new modes, which is analogous to the Autler-Townes splitting (ATS) observed in Floquet-driven atomic systems \cite{RevModPhys.89.011004,Eisert2015QuantumManyBody,PhysRevLett.99.220403,PhysRevLett.106.220402,PhysRevLett.119.123601}. This technique has demonstrated several novel capabilities, such as in-situ tuning of hybrid magnonic interactions \cite{Xu2020Dec_PRL} and selective magnon mode isolation \cite{Pishehvar2025Feb_PRAppl}. Such advancements present significant opportunities for further exploration and exploitation of hybrid magnonic systems.

However, previous demonstrations of the Floquet driving effect in hybrid magnonics have been relatively weak, hindering practical applications and the exploration of more intriguing physics. To address this issue, more efficient driving approaches are essential. In this work, we demonstrate that by leveraging resonance effects to amplify the Floquet drive, the Floquet magnonic interaction can be significantly enhanced. This enhancement is achieved by incorporating a capacitor into the driving coil to form LC resonances. With the improved Floquet drive, substantially enhanced ATS effects are observed in our system, allowing the observation of novel spectral shapes. Furthermore, we show that in the resonance-enhanced Floquet magnonic system, the Floquet drive can be generated from GHz signals while maintaining a sufficiently large amplitude. This also enables information-carrying signals in such magnonic systems to be used for control, leading to novel functionalities such as self-regulated attenuation and controlled transmission, which can be further implemented to integrated magnonic circuits for a wide variety of applications ranging from coherent information processing to magnon-based computing \cite{Wang2020MagnonicCoupler,Cocconcelli2025SelfBiasedMagnonic}.

\begin{figure}[bt]
    \centering
    \includegraphics[width=0.95\linewidth]{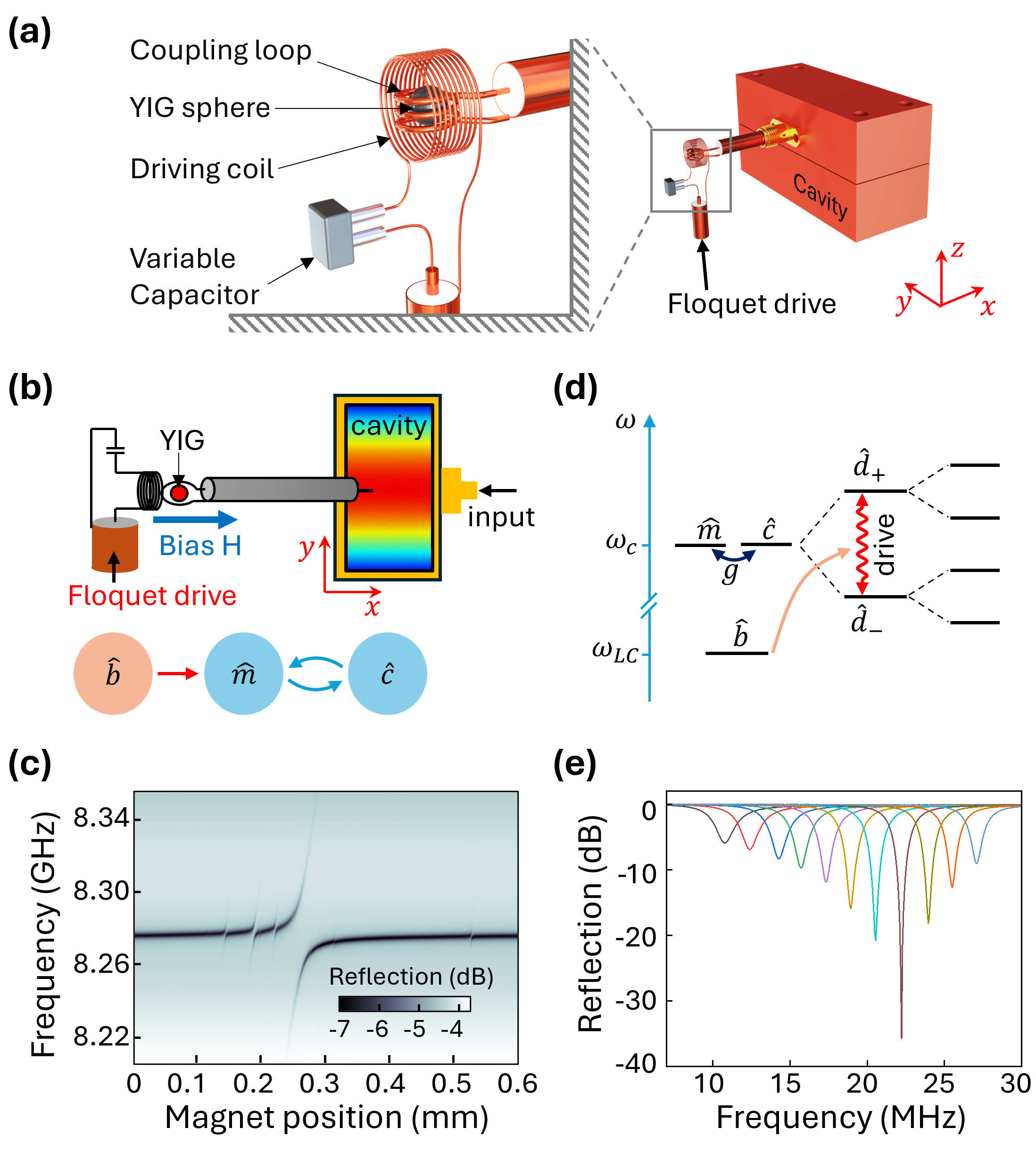}
    \caption{(a) Schematic drawing of the resonance-enhanced Floquet cavity electromagnonic device. A bias magnetic field is applied along x direction (parallel to the coupling coaxial cable). (b) Top view of the device schematics showing the interaction among the cavity photon ($\hat{c}$), magnon ($\hat{m}$), and LC resonator photon ($\hat{b}$). The heatmap inside the cavity shows the electric field distribution of the TE$_{011}$ mode inside the cavity. The input port is also used as the output for reflection measurements. (c) Measured cavity reflection spectrum measuring using a VNA with the bias magnetic field swept by varying the magnet position. (d) Equivalent energy diagram of our Floquet cavity electromagnonic system. (e) Measured reflection spectra of the LC resonator when the variable capacitor is tuned.}
    \label{fig1}
\end{figure}

\section{System Configuration} 

Figures\,\ref{fig1}(a) - (b) show the schematics of our device, which compromises a 3D microwave cavity and a highly polished YIG sphere. The 3D microwave cavity is machined from copper with a dimension of $9 \times 40 \times 20$ mm$^3$, supporting a TE$_{011}$ resonance at $\omega_c/2\pi=8.3$ GHz with a dissipation rate $\kappa_c/2\pi=2.5$ MHz. The YIG sphere has a diameter of 800 $\mu$m, which hosts a series of magnonic resonances. Different from previous demonstrations where the YIG spheres are placed inside the cavity, in our device the YIG sphere is located outside the cavity. A short semi-rigid coaxial cable couples the YIG sphere with cavity, with one end (open ended) inserted to the center of the cavity, while the other end has a shorted loop surrounding the YIG sphere. To support magnon excitation, a bias magnetic field is applied to the YIG sphere using a permanent magnet, along a direction parallel to the coaxial cable ($x$ direction). The frequency ($\omega_m$) of the magnon modes can be tuned by moving the magnet position, which varies the distance between the magnet and the YIG sphere, accordingly tuning the strength of the bias magnetic field to the YIG sphere.

Our device configuration allows for the separation of the YIG sphere with the cavity, simplifying the integration with the Floquet driving components; yet the coaxial cable allows efficient remote coupling between the magnon mode ($\hat{m}$) and the photon mode ($\hat{c}$). By sweeping the magnetic field, the cavity reflection spectra measured from another port of the cavity exhibit an anti-crossing feature, indicating the strong coupling between the fundamental magnon mode (the Kittel mode) and the cavity photon mode (the TE$_{011}$ resonance), as shown in Fig.\,\ref{fig1}(c). A coupling strength of $g_{cm}/2\pi=22$ MHz can be extracted from the spectrum, which can be further tuned by moving the insertion depth of the coaxial cable into the cavity. Two hybrid modes ($\hat{d}_\pm$) are generated through the magnon-photon coupling, as indicated Fig.\,\ref{fig1}(d).

To introduce the Floquet drive between the two hybrid modes, a driving coil is wrapped around the YIG sphere. With its direction parallel to the permanent magnet, a modulation to the bias field will be generated when the coil is fed by a sinusoidal signal through a coaxial cable. To improve the driving efficiency, an impedance-matching capacitor is connected to the coil in series, enabling the utilization of resonance photons to provide the Floquet drive, which leads to distinctive advantages compared with previous demonstrations. Using a variable capacitor, the resonance frequency of the LC resonator ($\omega_b/2\pi$) can be tuned from 10 MHz to 27 MHz, with the dissipation rate ($\kappa_b/2\pi$) varies between 2.5 and 1.5 MHz. The measured reflection spectra of the LC resonator in Fig.\,\ref{fig1}(e) show extinction ratios up to 36 dB, which far exceeds the maximum achievable extinction ratios (less than 1 dB, which corresponds to nearly total reflection) using non-resonant drives, indicating much enhanced energy delivery efficiency to the magnon mode.

\section{Theoretical Modeling} 

Our system can be described by a three-mode interaction model, involving the cavity photon $\hat{c}$, the magnon $\hat{m}$, and the LC resonator photon $\hat{b}$. The cavity photon and the magnon mode coherently interact with each other, and the LC resonator mode is parametrically coupled to the magnon mode in a way similar to the optomechanical interaction \cite{Aspelmeyer2014Dec,Barzanjeh2022Jan}. The system can be described by the Hamiltonian
\begin{equation}
    \begin{split}  
    \hat{H}/\hbar = & \omega_c \hat{c}^\dagger \hat{c} + \omega_m \hat{m}^\dagger \hat{m} + \omega_b \hat{b}^\dagger \hat{b} \\
    & +  g_{cm} (\hat{c}^\dagger \hat{m} + \hat{c} \hat{m}^\dagger) + g_{bm} \hat{m}^\dagger \hat{m}(\hat{b}+\hat{b}^\dagger) \\
    & + E \left( \hat{b} e^{i\omega_\mathrm{D} t} + \hat{b}^\dagger e^{-i\omega_\mathrm{D} t} \right) ,
    \end{split}
\end{equation}
where $\hat{c}$, $\hat{m}$, $\hat{b}$ ($\hat{c}^\dagger$, $\hat{m}^\dagger$, $\hat{b}^\dagger$) are the antihalation (creation) operator of cavity photon mode, magnon mode, and LC resonator photon mode, respectively, $\hbar$ is reduced Planck's constant, $g_{bm}$ is the single-photon coupling strength between the magnon and LC resonator photon, $\omega_\mathrm{D}$ is the drive frequency, and $E = \sqrt{\frac{\kappa_e P}{\hbar \omega_b}}$ is the drive amplitude with $P$ being the Floquet drive power at the device and $\kappa_e$ the external coupling rate of the LC resonator.

In our experiment, the classical drive is much stronger than the magnon-induced drive by several orders of magnitude ($E\gg g_{bm} \langle \hat{m}^\dagger  \hat{m} \rangle$). Therefore, $\hat{b}$ can be treated classically, with its dynamics determined by the quation of motion
\begin{equation}
    \dot{\hat{b}} = -\left( i\omega_b + \kappa_b \right) \hat{b} -i E e^{-i\omega_\mathrm{D} t} .
\end{equation}
At the steady state, we have
\begin{equation}
    \hat{b}(t) = \frac{-iE}{i(\omega_b-\omega_\mathrm{D}) + \kappa_b} e^{-i\omega_\mathrm{D} t} ,
\end{equation}
and the Hamiltonian for the cavity and magnon becomes
\begin{equation}
    \begin{split}
        \hat{H}/\hbar = ~& \omega_c \hat{c}^\dagger \hat{c} + \omega_m \hat{m}^\dagger \hat{m} + g_{cm}(\hat{c}^\dagger \hat{m} + \hat{c} \hat{m}^\dagger) \\
        & + \Omega \hat{m}^\dagger \hat{m} \cos(\omega_\mathrm{D} t) .
    \end{split}
    \label{Eq.Hamiltonian}
\end{equation}
Here
\begin{equation}
    \begin{split}
        \Omega =& \frac{g_{bm} E}{\sqrt{(\omega_b - \omega_\mathrm{D})^2 + \kappa_b^2}}
    \end{split}
\end{equation}
is the drive strength, and a phase delay in $\cos(\omega_\mathrm{D} t)$ has been ignored without loss of generality.

\begin{figure}[tb]
    \centering
    \includegraphics[width=0.98\linewidth]{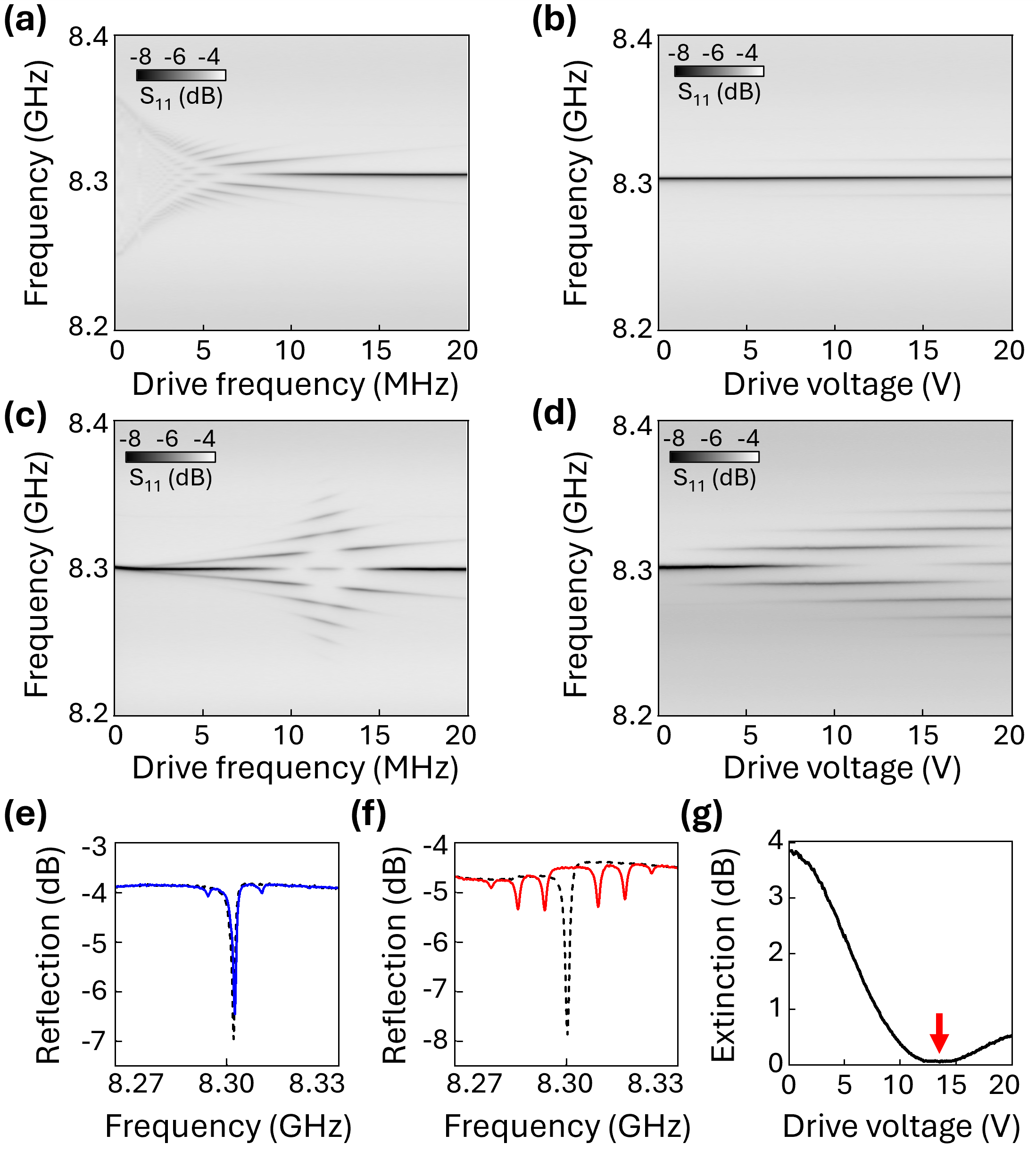}
    \caption{Measured reflection spectra of the magnon mode (without the cavity) under different Floquet drive conditions: (a) Without LC resonance, fixed modulation amplitude $V_\mathrm{D}= 20$ V, and varying modulation frequency; (b) Without LC resonance, fixed modulation frequency $\omega_\mathrm{D}/2\pi=12$ MHz, and varying modulation amplitudes; (c) With LC resonance, fixed modulation amplitude at 20 V, and varying modulation frequency; (d) With LC resonance, fixed modulation frequency at 12 MHz, and varying modulation amplitudes. (e)-(f) are the selected magnon reflection spectra without and with resonance enhancement, respectively, for a given modulation condition: $\omega_\mathrm{D}/2\pi=12$ MHz, $V_\mathrm{D}=20$ V. Dashed lines: magnon reflection spectrum without frequency modulation. (g) Magnon extinction ratio extracted from the reflection spectrum in (d) at frequency 8.3 GHz. Red arrow indicates where the center mode disappears (at $V_\mathrm{D}=14$ V).}
    \label{fig2}
\end{figure}

\section{Resonance-enhanced magnon modulation} 

To demonstrate the enhancement effect of the LC resonance on the Floquet drive, the response of the magnon mode is directly probed by measuring the reflection from the coupling loop. Specifically, the cavity is removed from the system, and the reflection spectrum is measured from the other end of the coaxial cable that couples to the YIG sphere. As a reference, the measurement results obtained without introducing the impedance-matching capacitor are plotted in Figs.\,\ref{fig2}(a)-(b).
Since the Floquet drive periodically modulates the bias magnetic field, the magnon mode is frequency modulated, exhibiting a series of sidebands. Note that in our Floquet magnonic system, the desired drive frequency is above 8 MHz to ensure clear separation of the hybrid modes, which usually have linewidths around 4 MHz. In this frequency range ($\omega_\mathrm{D}/2\pi >10$ MHz), only two weak sidebands (compared to the strong center mode) can be generated by the driving coil without the capacitor, even with large driving amplitudes up to 20 V (peak-to-peak value, as used throughout this work).

When a capacitor is connected in series to the coil, the spectrum exhibits a total of eight enhanced sidebands, as well as a highly suppressed center mode, near the LC resonance frequency ($\omega_b/2\pi=12$ MHz) at an amplitude of 20 V, as shown in Fig.\,\ref{fig2}(c). Outside the LC resonance, i.e., when $\omega_\mathrm{D}/2\pi<10$ MHz or $\omega_\mathrm{D}/2\pi>14$ MHz, only two sidebands can be generated. Figure\,\ref{fig2}(d) plots the magnon sidebands generated at different drive amplitudes by a 12 MHz Floquet drive. When the drive amplitude is below 5V, there exist two sidebands whose strength gradually increases with the drive amplitude. As the drive amplitude increases, more high-order sidebands start to appear. At a drive voltage of 18 V, the total number of visible sidebands increases to eight. The curves in Figs.\,\ref{fig2}(e)-(f) show a more direct comparison of the LC-resonance enhancement effect for the Floquet drive. For the same Floquet drive with an amplitude of 12 V and frequency $\omega_\mathrm{D}=12$ MHz, the sideband generated without the LC resonance enhancement (blue curve) only has an extinction ratio of 0.2 dB with a strong center mode (extinction ratio: 3 dB); In contrast, with the LC resonance (red curve), the first two sidebands (among a total of six sidebands) exhibit an extinction ratio of about 1 dB while the center mode is completely suppressed.

It is worth noting that the dependence of the center mode and the sideband strength on the drive amplitude is not monotonic. As an example, when the drive amplitude increases from 10 V to 20 V, the strength of the first sideband slowly decreases, while the center mode first decreases towards a minimum at 14 V (indicated by the red arrow) and then increases again. This dependence can be simply explained by the frequency modulation model for the magnon mode, whose dynamics can be described by the equation below
\begin{equation}
         \dot{\hat{m}}  =  -i (\omega_m + \Omega \cos (\omega_D t)) \hat{m}.
\end{equation}
This equation of motion is based on the Hamiltonian in Eq.\,(\ref{Eq.Hamiltonian}), with only the magnon mode considered. The solution of the equation is 
\begin{equation}
    \hat{m}(t) = e^{-i\omega_m t - i\Omega/\omega_D \sin (\omega_D t)}.
\end{equation}
This leads to the output signal of the magnon mode
\begin{equation}
    \hat{m}(t)+\hat{m}^\dagger (t) = \cos \left(\omega_m t + \frac{\Omega}{\omega_D} \sin (\omega_D t) \right),
\end{equation}
which is frequency modulated. In this model, the dependence of the sideband amplitude on the drive strength follows the Bessel function of the first kind: $A_n \propto J_n(\frac{\Omega}{\omega_\mathrm{D}})$, where $n$ is the order of the sideband. For the carrier ($n=0$), the first root of the Bessel function $J_0(\frac{\Omega}{\omega_\mathrm{D}})$ occurs at $\Omega/\omega_\mathrm{D}=2.4$.  Given a Floquet drive of $\omega_\mathrm{D}/2\pi=12$ MHz, this leads to $\Omega/2\pi=28.8$ MHz, which is the maximum frequency shift induced by the frequency modulation. It corresponds to a Floquet-drive-induced magnetic field $h=\Omega/\gamma\approx 10$ Oe, where $\gamma/2\pi=2.8$ MHz/Oe is the gyromagnetic ratio. Based on Fig.\ref{fig2}(g), this occurs at a drive amplitude of 14 V where the extinction ratio of the center mode reaches the minimum.

\begin{figure}[tb]
    \centering
    \includegraphics[width=0.98\linewidth]{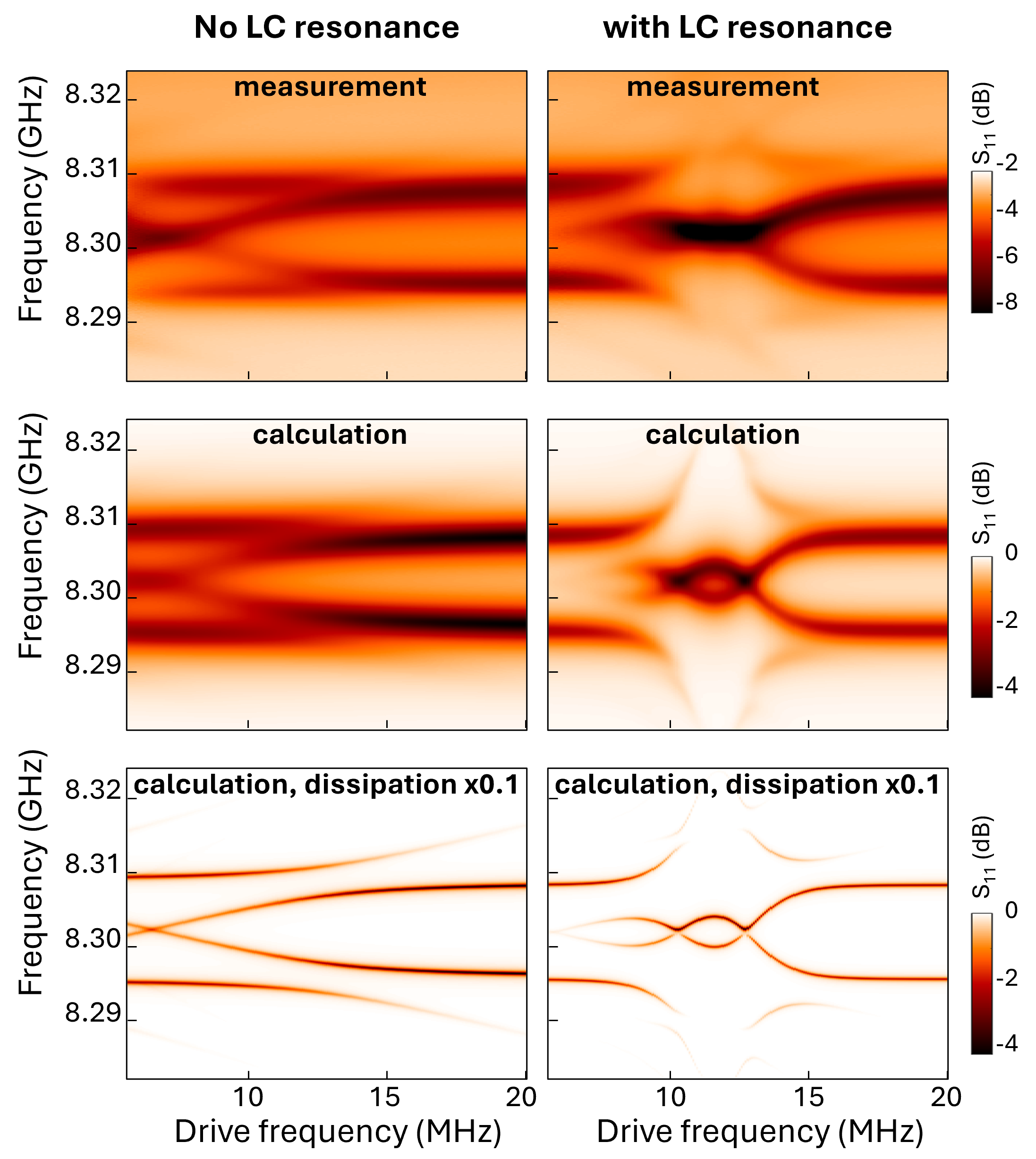}
    \caption{Cavity reflection spectra obtained from measurement (top row), calculation using realistic dissipation (middle row), and calculation with 10-times reduced dissipation rates (bottom row), respectively. Left column is obtained without the LC resonance, while the right column with the LC resonance. The Floquet drive peak-to-peak amplitude is 20 V; the LC resonance frequency $\omega_b$ and the hybrid mode frequency difference $\Delta$ are both set to $2\pi\times 12$ MHz.}
    \label{fig3}
\end{figure}

\begin{figure*}[tb]
    \centering
    \includegraphics[width=0.98\linewidth]{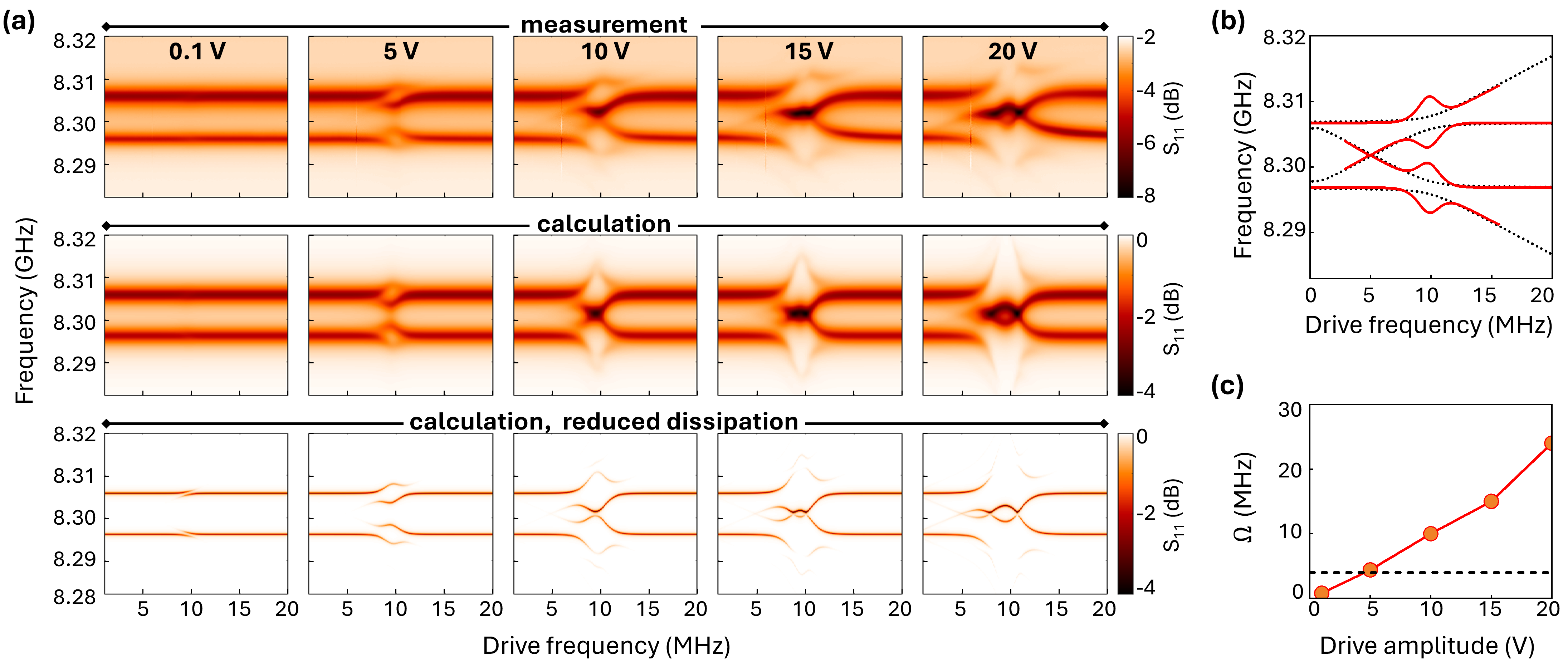}
    \caption{(a) Cavity reflection spectra with LC resonance enhancement at different driving amplitudes. Top row: measurement; middle row: calculation using realistic dissipation rates; bottom row: calculation with $10\times$ reduced dissipation rates. The LC resonance frequency $\omega_b$ and the hybrid mode frequency difference $\Delta$ are both set to $2\pi\times 10$ MHz. An eye pattern is visible at 10 MHz for each hybrid mode at 5-V drive amplitude, which keeps widening as the drive amplitude increases. (b) Schematic illustration of the increased ATS near the LC resonance frequency (solid red), as compared with the anti-crossing feature under conventional Floquet drive (dotted black). (c) Floquet drive strength $\Omega$ extracted from the numerical fitting as a function of the drive signal amplitude (red). Dashed black curve: the maximum achievable Floquet drive strength $\Omega$ extracted for the case without LC resonance enhancement.}
    \label{fig4}
\end{figure*}

\section{Resonance-enhanced Floquet magnonics} 

Compared with previous demonstration of Floquet magnonics, the ATS generated in our resonance-enhanced Floquet magnonic system exhibits drastically different spectral features. This is clearly shown by comparing the cavity reflection spectra measured with and without the LC resonance enhancement effect for the same Floquet drive (20 V), as plotted in Fig.\,\ref{fig3}. In these measurements, the magnon mode is first tuned to be on resonance with the cavity mode by varying the bias magnetic field, which leads to the formation of two hybrid modes $\hat{d}_\pm$ separated by $\Delta =2g_{cm}=2\pi\times 12$ MHz in frequency. When the Floquet drive is applied, the two hybrid modes couple with each other and ATS occurs when the drive frequency matches the mode spacing ($\omega_\mathrm{D}=\Delta$). Without the LC resonance enhancement (left column in Fig.\,\ref{fig3}), the measured spectrum exhibits an anti-crossing feature when the drive frequency is swept across the condition $\omega_\mathrm{D}=\Delta$, which is similar to our previous demonstrations \cite{Xu2020Dec_PRL} and agrees well with the theoretical calculation. However, the ATS is very small because of the poor driving efficiency, which is due to the impedance mismatch of the coil. From the spectrum, an ATS of $2\pi \times 3.9$ MHz is extracted, which is even less than the linewidth (full width at half maximum) of the hybrid modes ($2\kappa_\pm/2\pi=4$ MHz). As a result, the anti-crossing feature is not very visible. To better reveal this spectral feature, another calculation is performed using dissipation rates that are artificially reduced by 10 times, as shown in the bottom row of Fig.\,\ref{fig3}. As a comparison, when the LC resonance is introduced with a frequency matching the hybrid mode frequency difference ($\omega_b=\Delta$), the ATS is drastically enhanced, as shown in the right column of Fig.\,\ref{fig3}. The calculation results for this case reveal more intriguing features, where the ATS even exceeds $\Delta$, causing the additional mode crossing features near the middle of the spectrum.

To study the effect of the drive amplitude on the resonance-enhanced Floquet magnonics, a Floquet drive with varying amplitudes (from 0.1 V to 20 V) is applied to a device with $\Delta/2\pi= 10$ MHz, and the measured reflection spectra are plotted in the first row of Fig.\,\ref{fig4}. In this experiment, the LC resonance is tuned to match the hybrid mode frequency difference. When the drive amplitude is low (e.g., below 10 V), an eye pattern can be observed on each hybrid mode when the drive frequency matches the LC resonance and the hybrid mode frequency difference, which is drastically different from the conventional anti-crossing features. This can be explained by the fact that the enhancement effect only takes place within the finite linewidth (4 MHz) of the LC resonance, which locally increases the splitting of the anti-crossing feature and leads to the eye patterns [Fig.\,\ref{fig4}(b)]. As the drive amplitude increases, the eye pattern gradually widens. When the drive amplitude reaches 10 V, the upper branch of the bottom eye pattern meets the lower branch of the top eye pattern. When the drive amplitude increases beyond 10 V, these two branches will cross each other, leading to the mode inversion near the LC resonance frequency.

Our experimental results agree well with the theoretical calculation [middle row in Fig.\,\ref{fig4}(a)], which utilize the actual linewidths for the hybrid modes extracted from numerical fitting of the experimental data. We also performed another calculation with reduced linewidths (by 10 times) to better illustrate the novel mode distribution induced by the resonance-enhanced Floquet drive, including the mode merging and mode inversion, and the results are plotted in the bottom row of Fig.\,\ref{fig4}(a). The drive strength $\Omega$ for each hybrid mode is extracted from our numerical model and plotted in Fig.\,\ref{fig4}(c) (red dots) as a function of the drive amplitude. As a comparison, the maximum drive strength induced by the same Floquet drives (up to 20 V) without the LC resonance enhancement is also plotted (dashed horizontal line), showing an enhancement of around six times when a LC resonance is adopted. In addition, our measurements show that such enhancement effect not only works when the LC resonance is tuned to match the hybrid mode spacing. When LC resonance is tuned off resonance, enhanced ATS can still be induced, which shows the effectiveness of the enhancement effect induced by the LC resonance (see Fig.\,\ref{figSM1} in Appendix A).

\section{Two-tone drive} 

With the resonance-enhanced drive, existing Floquet magnonic functionalities, such as tunable coherent coupling \cite{Xu2020Dec_PRL} and on-demand magnon mode isolation \cite{Pishehvar2025Feb_PRAppl}, can be significantly enhanced. In addition, it will allow more advanced functionalities that far exceed previous magnonic systems. For instance, one limitation of the previous demonstrations is that the parametric control of the system cannot exceed MHz frequencies. Therefore, the GHz output signal for a given device, either in the form of magnons or photons, can hardly be used as the control signal for the next (or the same) device, limiting the possibility of device cascading or feedback control. With the LC-resonance enhancement, it is possible to use GHz signals to control the dynamics of the system.

\begin{figure}[tb]
    \centering
    \includegraphics[width=0.98\linewidth]{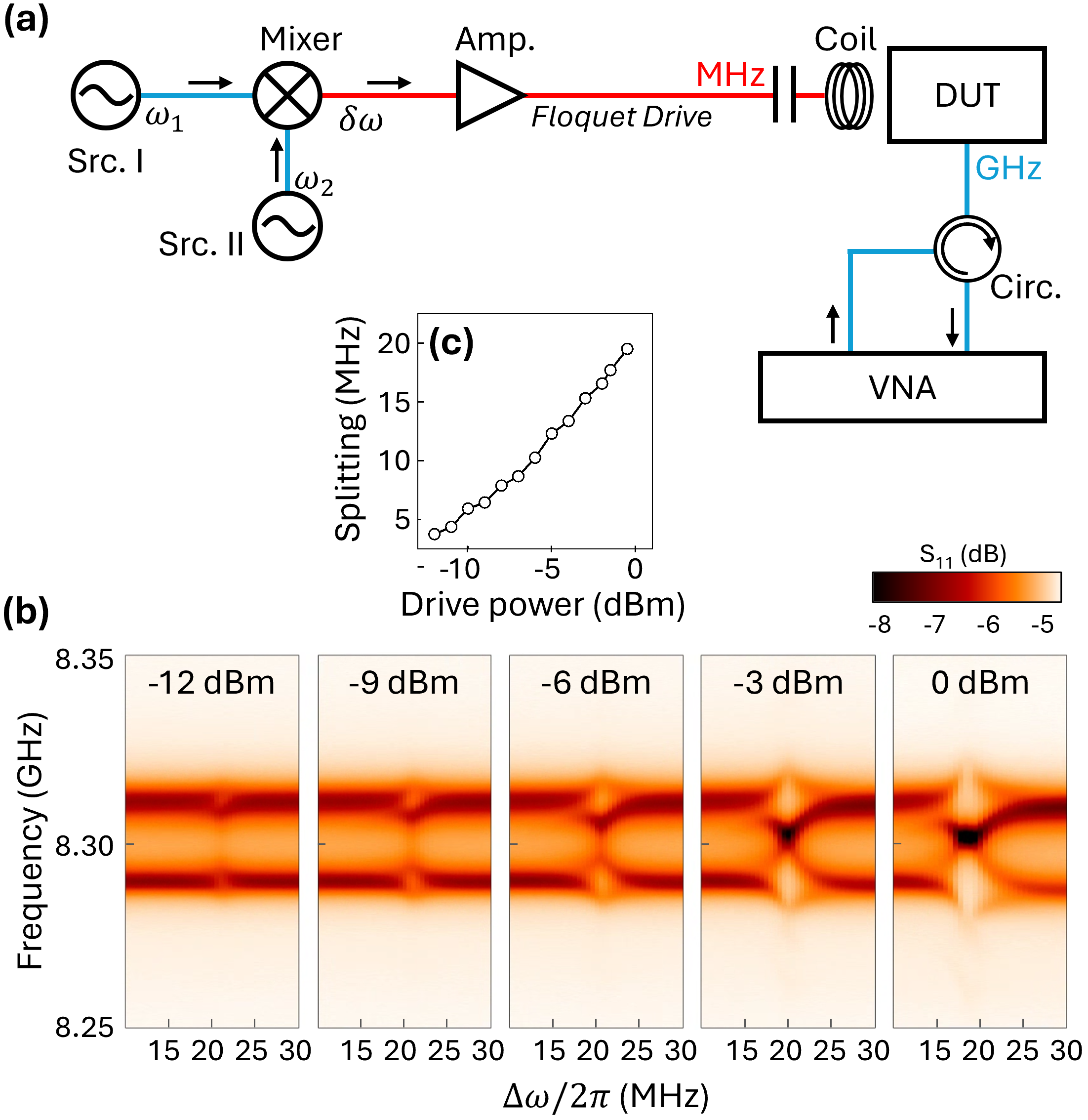}
    \caption{(a) Schematics of the Floquet drive experiment using two microwave tones. The drive comes from the mixing of two GHz signals (Src. I and Src. II). The YIG sphere coupled to the cavity is not shown for clarity. (b) Measured reflection spectra at different powers for Source II, with Source I power fixed at 0 dBm. (c) Extracted ATS as a function of drive power (from Source II). Src.: source; LPF: low-pass filter; Amp.: amplifier; Cap.: capacitor; Circ.: circulator.}
    \label{fig5}
\end{figure}

Figure\,\ref{fig5}(a) shows the schematics of our experimental setup to demonstrate such capability. Two GHz signal sources (at frequencies $\omega_1$ and $\omega_2$) are mixed in a microwave mixer to generate a signal at frequency $\delta\omega=|\omega_1-\omega_2|$. By controlling the detuning of the two GHz signals, $\delta\omega$ can be tuned to match the MHz LC resonance ($\delta\omega=\omega_b$) and thus the mixed signal can be used as an efficient Floquet drive. To further boost its strength, an amplifier is used before the signal is sent to the LC resonator. The device response is monitored by measuring the cavity reflection spectrum using a VNA, similar to Figs.\,\ref{fig3}-\ref{fig4}. The external bias magnetic field is tuned to match the magnon frequency with the cavity resonance, which generates two hybrid modes (at $\omega_+/2\pi=8.3105$ GHz and $\omega_-/2\pi=8.2895$ GHz) separated by $\Delta/2\pi=21$ MHz. The LC resonance is tuned to match the hybrid mode frequency difference ($\omega_b=\Delta$), and the detuning frequency $\delta\omega$ is swept across the LC resonance frequency. Note that the frequencies of the two signals that are mixed to generate the Floquet drive can be completely different from the frequencies of the two hybrid modes ($\omega_{1,2}\ne \omega_\pm$). In fact, they can be selected arbitarily, as long as their difference matches the LC resonance frequency. In our experiment, Source I is fixed at 8.3 GHz, while Source II is swept from 8.311 to 8.330 GHz, corresponding to a range of 11 
to 30 MHz for $\delta\omega/2\pi$. 

Figure\,\ref{fig5}(b) shows the measured cavity reflection spectra, which exhibit similar eye-patterns as obtained in the single-drive case in Fig.\,\ref{fig4}. The measurement is performed for a series of power levels ranging from -12 to 0 dBm for Source II, with Source I fixed at 0 dBm power. When the the power of Source II increases, The eye-pattern gradually widens, similar to the spectra obtained using a single MHz drive. Mode merging and inversion can be observed for power level of -3 dBm and 0 dBm, respectively. The size of the eye-pattern is extracted and plotted as a function of Source II power in Fig.\,\ref{fig5}(c), showing a maximum splitting of 20 MHz for each hybrid mode. In our experiment, this is achieved when both sources are set to 0 dBm, which can be further increased with larger input powers.

\begin{figure}[tb]
    \centering
    \includegraphics[width=0.98\linewidth]{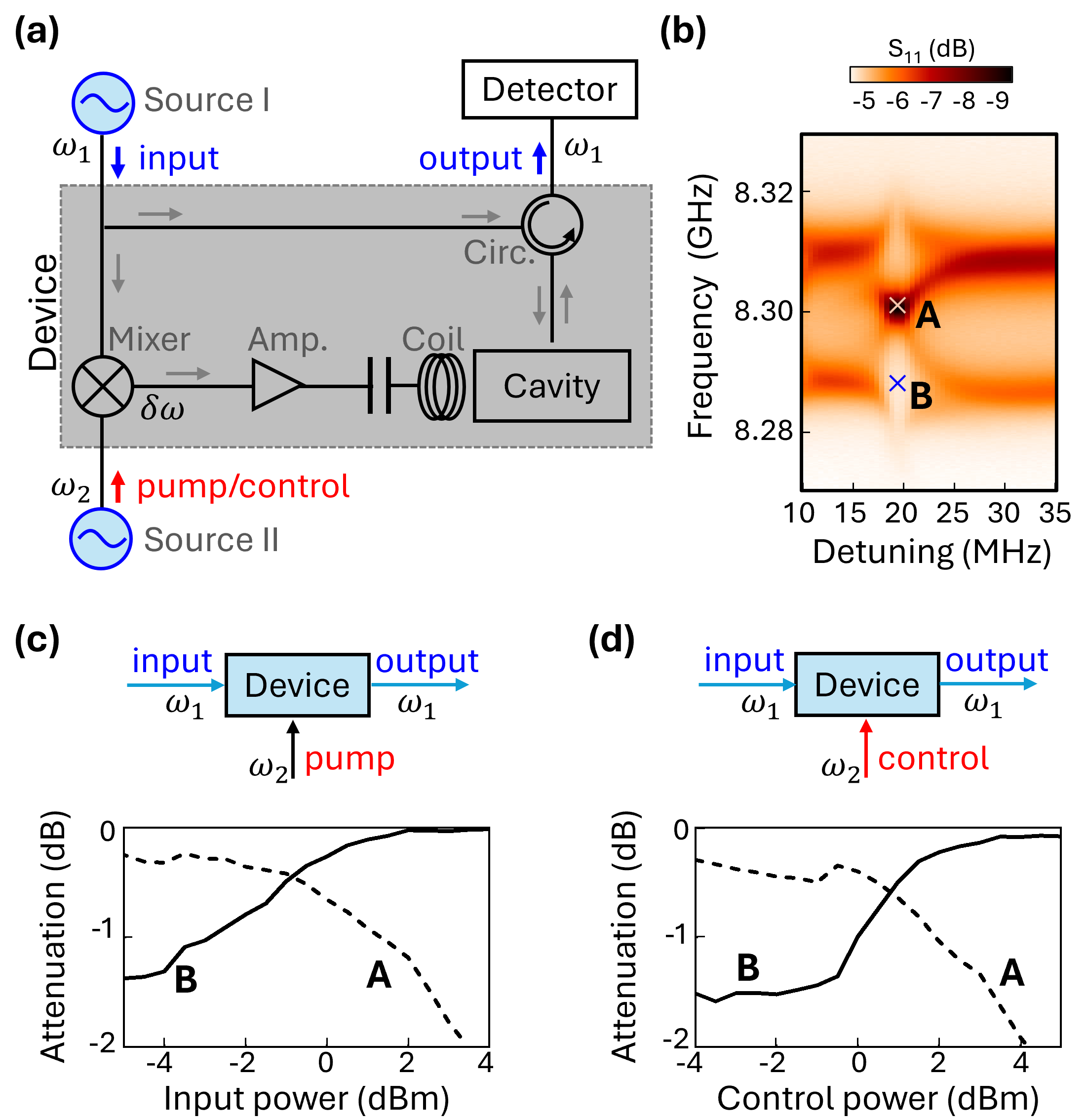}
    \caption{(a) Schematics of the experimental setup for demonstrating the Floquet gate operation. (b) The cavity reflection spectrum measured using VNA, indicating the two Operating Points A and B for the gate operation in (a). (c) Device attenuation as a function of the input power (with fixed pump power at 0 dBm) for both operating points in (b), showing the self-regulating attenuation functionality. (d) Device attenuation as a function of the control signal power (with fixed input power at 0 dBm) for both Operating Points A and B, showing the controlled transmission functionality.}
    \label{fig6}
\end{figure}

\section{Floquet gate operation} 

To demonstrate the application potential of the two-tone Floquet drive method, we used the setup schematically shown in Fig.\,\ref{fig6}(a) to implement two practical functionalities. In this configuration, the cavity and magnon modes are configured in the same way as in Fig.\,\ref{fig5}, exhibiting a cavity reflection spectrum as shown in Fig.\,\ref{fig6}(b). The LC resonance frequency is tuned to match the hybrid mode frequency difference, i.e., $\omega_b=\Delta$. In these application scenarios, a sinusoidal signal is sent to the device, and the device output (reflection of the input signal) is controlled by another signal (the control signal), or by the input signal itself (assisted by a pump signal). Two signal sources are used, one at $\omega_1$ as the input signal, while the other at $\omega_2=\omega_1+\omega_b$ as the control or the pump signal. The device dynamics is controlled by the two-tone Floquet drive, which is generated by mixing the signals from both sources. The mixer output is amplified and then sent to the LC resonator to achieve enhanced and controllable Floquet magnonic interactions. 

Figure\,\ref{fig6}(c) shows the first functionality: self-regulating attenuation. With the presence of a pump signal (from Source II. Frequency: $\omega_2$. Power: 0 dBm), the input signal (from Source I. Frequency: $\omega_1$) will regulate its own output attenuation. Depending on the selection of operating points [A or B in Fig.\,\ref{fig6}(b)], the output attenuation is increased or decreased as the input power increases. For Operating Point A ($\omega_1/2\pi=8.289$ GHz, $\omega_2/2\pi=8.309$ GHz), when the input power is very low, the mixer output and accordingly the Floquet drive is weak, leading to nearly closed eye-patterns for each hybrid mode. As a result, the mode merging at Operating Point A will disappear, leading to high output and accordingly low attenuation. As the input power increases, the mixer output increases, leading to increased Floquet drive and accordingly wider eye-patterns. Consequently, the mode merging occurs at Operating Point A, causing the device output to decrease and accordingly exhibiting high attenuation. In other words, low (high) input signal power is always accompanied by low (high) attenuation, thereby behaving as a self-regulating attenuator, which in practice can be used for power limiting functions. If Operating Point B is selected, then the input signal with low (high) amplitude will experience high (low) attenuation, i.e., weaker signals get damped further while stronger signals experiences less damping, which can be used as a dynamic range extender.

Figure\,\ref{fig6}(d) plots the second functionality: controlled transmission. In this demonstration, by sweeping the power of the control signal at $\omega_2$, the device output at Operating Points A and B are plotted as dashed and solid lines, respectively, which exhibit similar trends as the self-regulating attenuation plotted in Fig.\,\ref{fig6}(c). Compared with the self-regulating attenuation, here the output signal is not controlled by the input signal strength, but instead by the amplitude of another control signal. Considering both the input and control signals operate in the same GHz frequency range, such a tuning mechanism allows the interconnect of multiple Floquet hybrid magnonic devices operating at similar frequencies, where the output from one device can control the behavior of the next device, allowing the implementation of complex logic operations. This has important implications for the development of practical applications in magnonic logic and related areas. Our operation principle, although demonstrated using bulky discrete devices, can be readily applied to integrated magnonic circuits and contribute to the advancement of magnonic computing technology.

\section{Appendix A: Off-Resonance LC drive}
\begin{figure}[tb]
    \centering
    \includegraphics[width=0.98\linewidth]{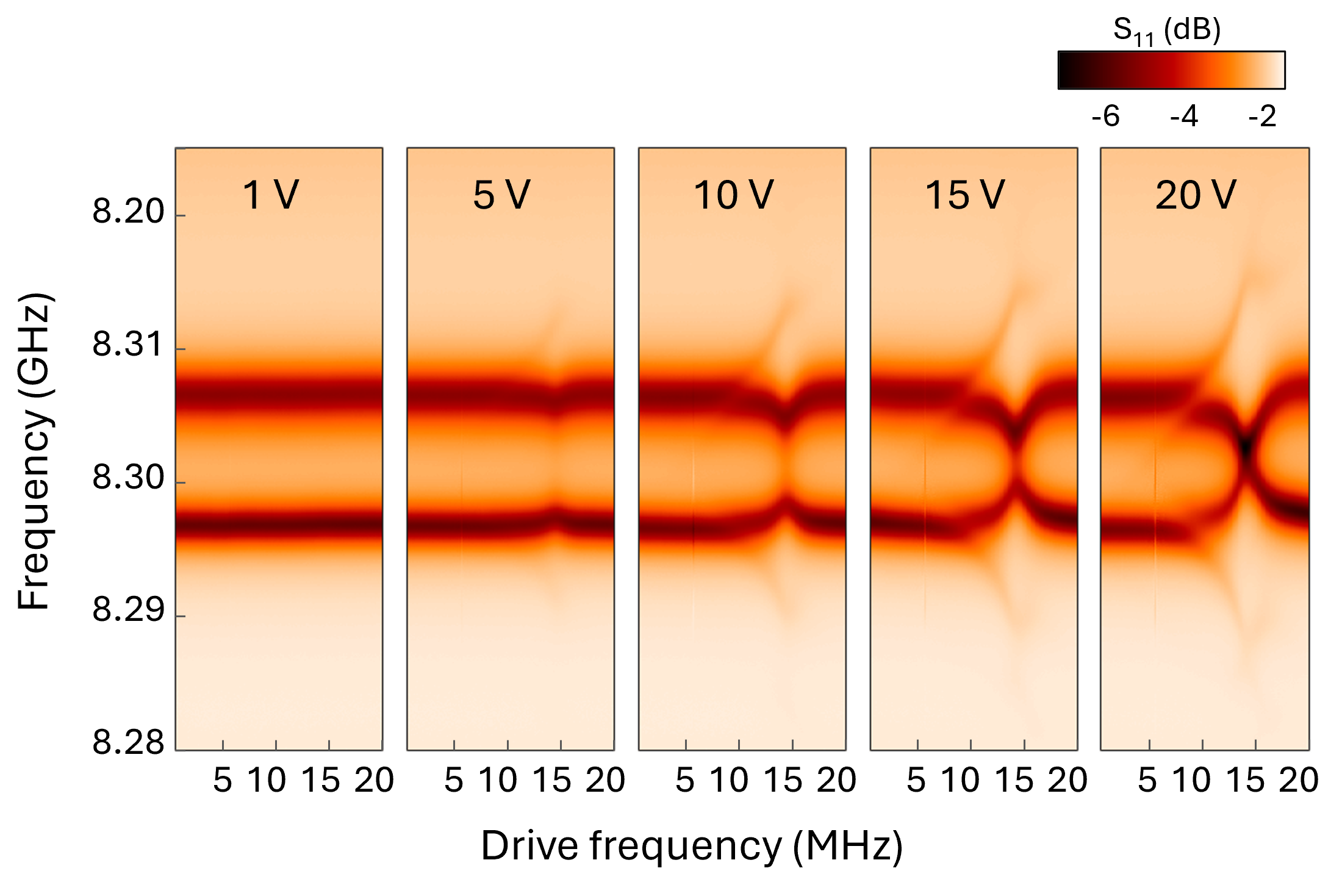}
    \caption{Measured spectra at different amplitudes of a single MHz drive. The LC resonance frequency is tuned to $\omega_b/2\pi= 15$ MHz, which differs from the hybrid mode frequency difference $\Delta/2\pi= 10$ MHz.}
    \label{figSM1}
\end{figure}

Because of the strong enhancement effect of the LC resonance, strong ATS can still be observed even when the LC resonance is offset from the  frequency difference of the two hybrid modes $\hat{d}_\pm$. Figure\,\ref{figSM1} shows the measured cavity reflection spectra for a single MHz Floquet drive at different drive amplitudes ranging from 1 V to 20 V. The hybrid mode frequency difference is set at $\Delta/2\pi=10$ MHz, but the LC resonance frequency is tuned to $\delta\omega/2\pi=15$ MHz. As the Floquet drive frequency is swept, an anti-crossing feature is observed as $\delta\omega/2\pi=10$ MHz. However, more prominent ATS is observed at $\delta\omega/2\pi=15$ MHz. In particular, at a 20-V drive amplitude, mode merging is observed. This indicates that a Floquet drive at 15 MHz induces strong interaction between the two hybrid modes $\hat{d}_\pm$ that are separated by 10 MHz, which is attributed to the significant resonance-induced enhancement effect to the off-resonance drive. 

\section{Acknowledgments}
\begin{acknowledgments}
X.Z. acknowledges support from NSF (2337713).  L.J. acknowledge support from the ARO(W911NF-23-1-0077), ARO MURI (W911NF-21-1-0325), AFOSR MURI (FA9550-21-1-0209, FA9550-23-1-0338), NSF (OMA-2137642, OSI-2326767, CCF-2312755, OSI-2426975), and Packard Foundation (2020-71479). J.M.J acknowledges support from AFOSR (FA9550-23-1-0254). 
\end{acknowledgments}

\bibliography{maintext}

\end{document}